\documentclass{JAC2003}

\usepackage{graphicx}
\usepackage{booktabs}

\begin{document}
\title{\rightline{\vspace{-.1in}\small\rm IIT-CAPP-13-2}
\rightline{\vspace{-.1in}\small\rm  MICE-CONF-GEN-415}
\rightline{\vspace{-.1in}\small\rm FERMILAB-CONF-13-172-APC}
\vspace{-.4in}
MUON COOLING, MUON COLLIDERS,\\ AND THE MICE EXPERIMENT\thanks{To appear in Proc. COOL'13 Workshop, M\"{u}rren, Switzerland, 10--14 June 2013.}\thanks{ Work supported by the U.S. DOE and NSF.}\vspace{-.15in}}

\author{Daniel M. Kaplan\thanks{kaplan@iit.edu}, Illinois Institute of Technology, Chicago, IL 60616, USA\\
on behalf of the MAP and MICE collaborations}\vspace{0.15in}

\maketitle

\begin{abstract}
   Muon colliders and neutrino factories are attractive options for future
facilities aimed at achieving the highest lepton-antilepton collision
energies and precision measurements of parameters of the Higgs
boson and the neutrino mixing matrix. The performance and cost of
these depend on how well a beam of muons can be cooled.
Recent progress in muon cooling design studies and prototype tests
nourishes the hope that such facilities can be built during the coming
decade. The status of the key technologies and their various
demonstration experiments is summarized.
\end{abstract}

\section{Muon Colliders and Neutrino Factories}
\begin{figure}[t]
\vspace{0.1in}\includegraphics[width=\linewidth]{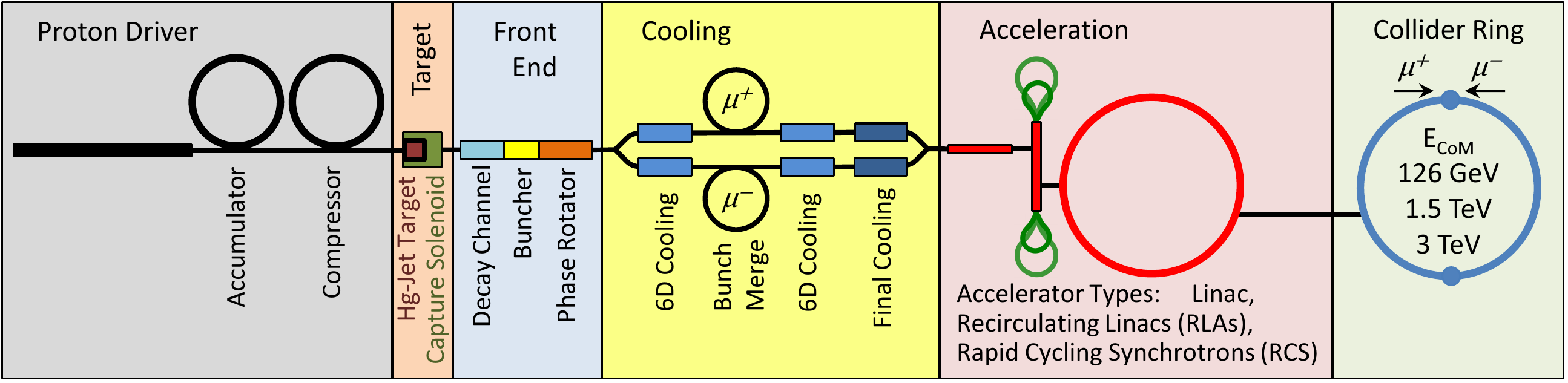}\\[0.1in]
\includegraphics[width=\linewidth]{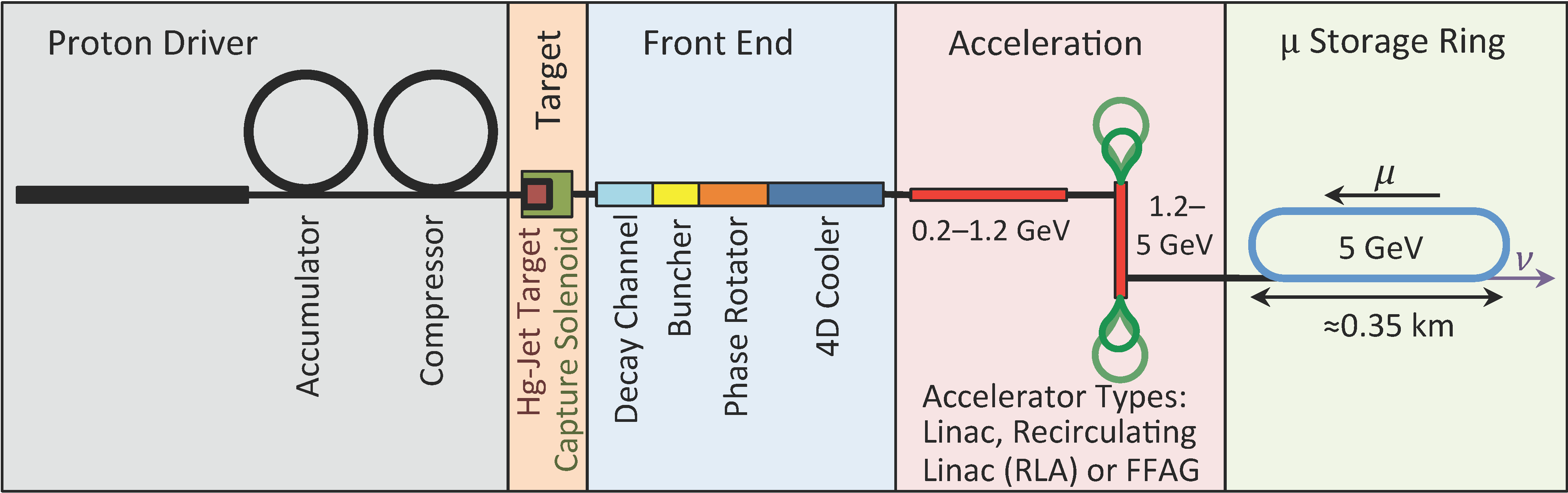}
\caption{(top) Muon collider and (bottom) neutrino factory schematic diagrams.} 
\label{fig:MC-NF} 
\end{figure}

\begin{figure}[h]
\centerline{\includegraphics[width=\linewidth]{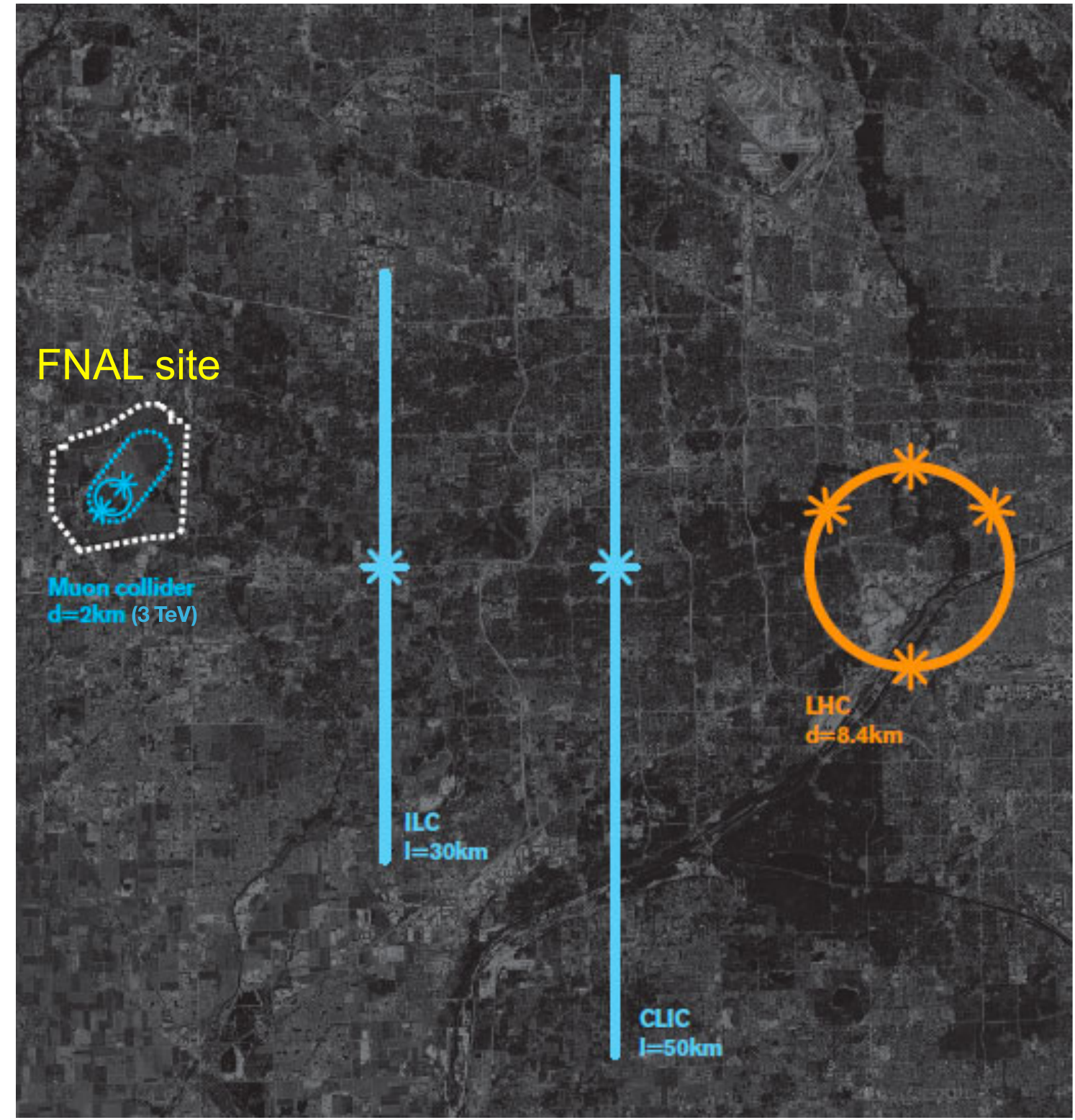}}
\caption{Collider sizes compared with FNAL
site. A muon collider with $\sqrt{s}>3$\,TeV fits on existing
sites.}\label{fig:sizes}
\end{figure}

\begin{figure}[h]
\includegraphics[width=.47\linewidth]{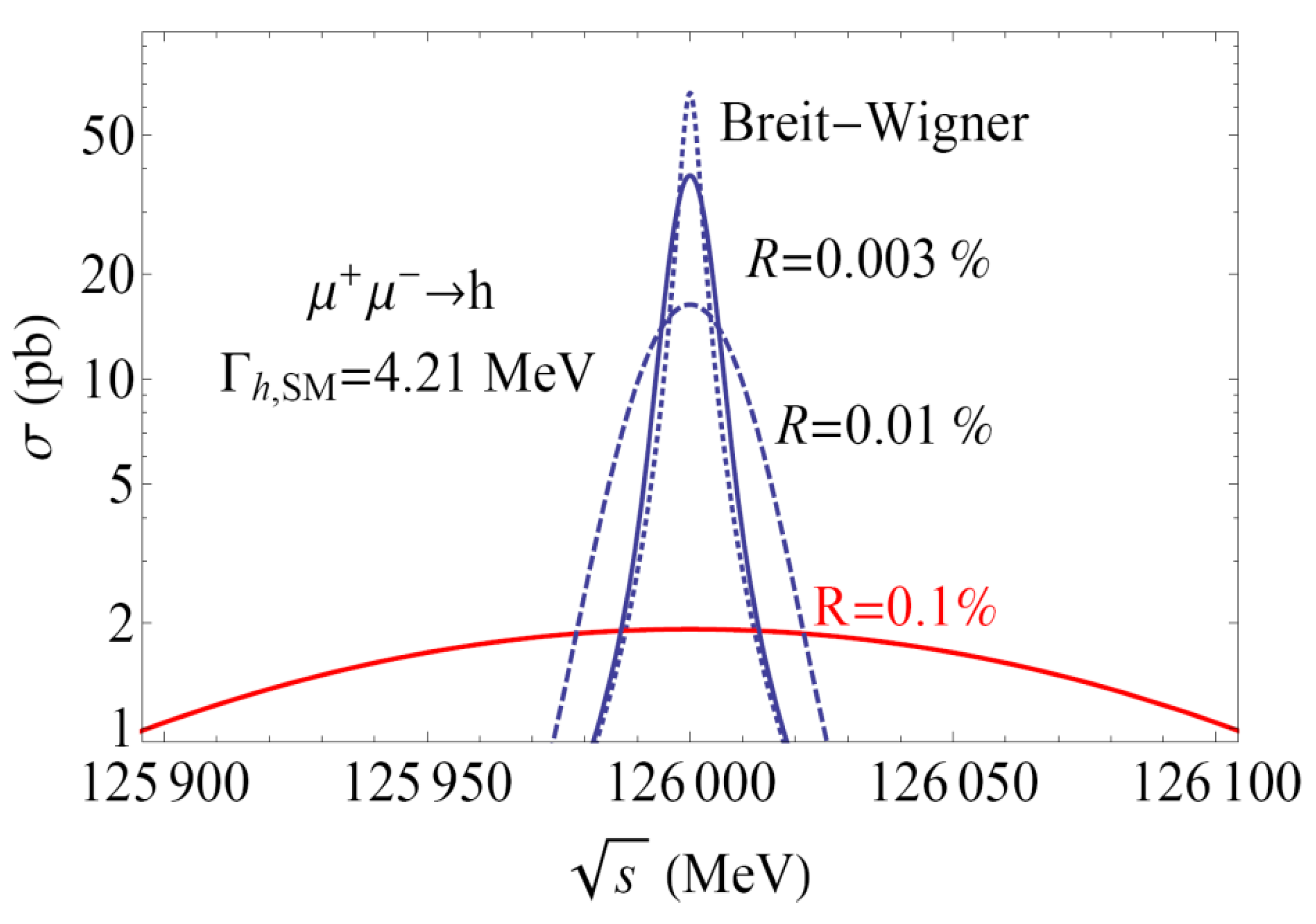}
\includegraphics[width=.52\linewidth]{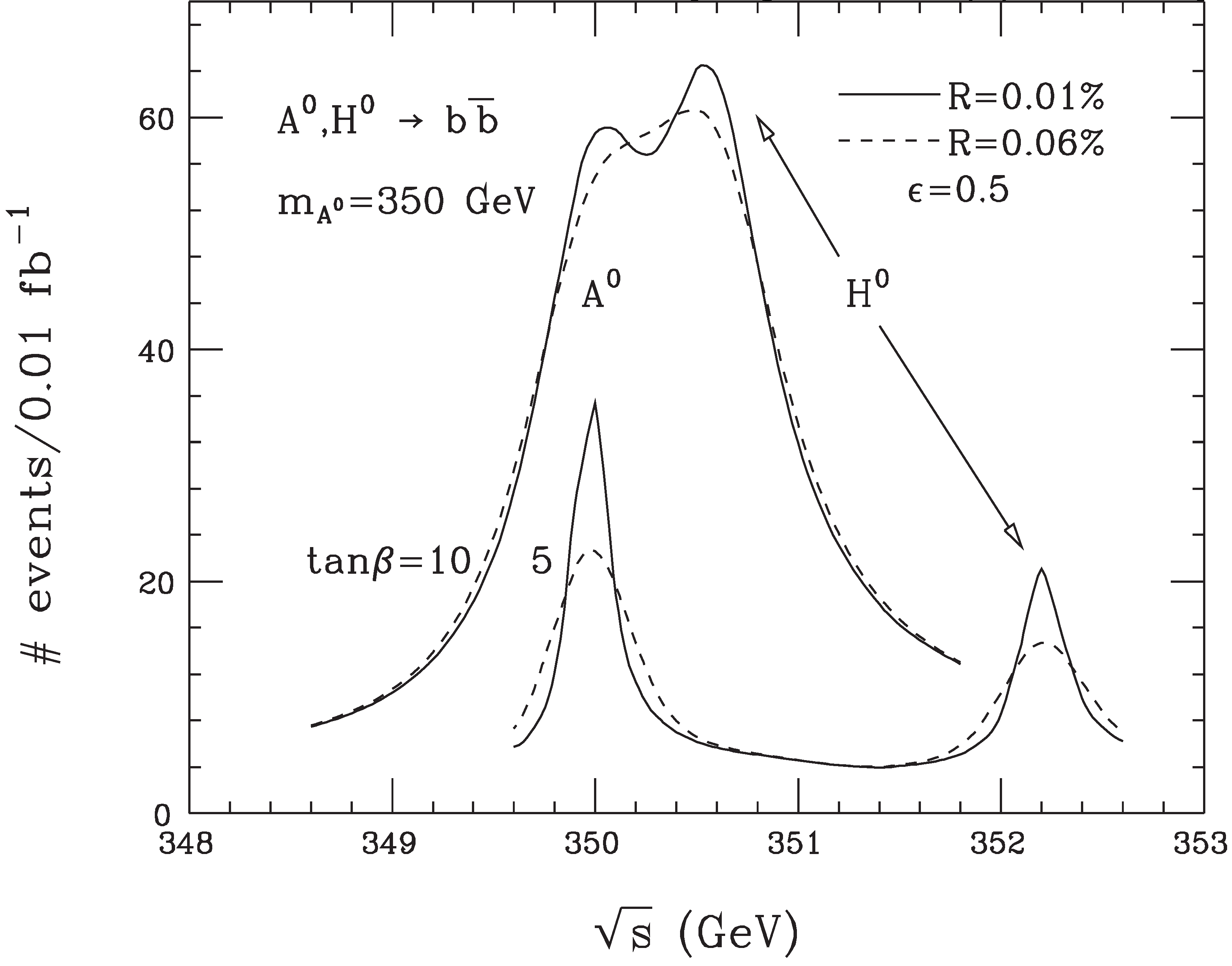}
\vspace{-.1in}\caption{(left) Standard Model Higgs line shape compared with three scenarios for muon collider energy resolution~\protect\cite{HF2012}; (right) resolving scalar and pseudoscalar supersymmetric Higgs partners for two possible values of the supersymmetric parameter $\tan{\beta}$~\protect\cite{Barger-Snowmass}.}\label{fig:Higgs}
\end{figure}

Discussed since the 1960s~\cite{Tikhonin-Budker,MC}, muon colliders (Fig.~\ref{fig:MC-NF}) are now reaching the threshold at which their construction can be realistically contemplated. Their interest stems from the important advantages over electrons that muons confer for high-energy lepton colliders:
suppression of radiative processes  by the 200-times greater mass of the muon, enabling the use of storage rings and
recirculating accelerators, and of
 ``beamstrahlung" interactions, which limit 
$e^+e^-$-collider luminosity as energy  increases~\cite{Palmer-Gallardo}.  The smaller size of a muon collider (Fig.~\ref{fig:sizes}) eases the siting issues and suggests that the cost will be less as well.
Furthermore, the muon/electron cross-section ratio for $s$-channel annihilation to Higgs bosons, $({m_\mu}/{m_e})^2=4.3\times10^4$, gives the muon collider unique access to precision Higgs measurements~\cite{HF2012,Bargeretal,Barger-Snowmass,HF}. For example, at the $\approx$\,126\,GeV/$c^2$ mass measured by ATLAS and CMS~\cite{LHC-Higgs}, only a muon collider can directly the observe the (4\,MeV) width and lineshape of a Standard Model Higgs boson~\cite{HF2012}  (see Fig.~\ref{fig:Higgs}). Furthermore, should the Higgs have closely spaced supersymmetric partner states at higher mass, only a muon collider has the mass resolution required to distinguish them. (The same argument applies as well to closely spaced scalar states in any other new-physics scenario.)

The neutrino factory (Fig.~\ref{fig:MC-NF}) is a newer idea~\cite{Geer}. A muon storage ring is  an ideal source for long-baseline neutrino-oscillation experiments: via $\mu^-\to e^-\nu_\mu{\overline \nu_e}$ and $\mu^+\to e^+{\overline \nu_\mu}\nu_e$, it can provide collimated, high-energy neutrino 
beams with well-understood composition and properties.  The clean identification of final-state muons in far detectors  
enables low-background appearance measurements using $\nu_e$ and $\overline \nu_e$ beams. Distinguishing oscillated from non-oscillated events requires a magnetized detector: if $\mu^-$ are stored in the ring, the oscillated events contain $\mu^+$, and vice versa if $\mu^+$ are stored. Now that a non-zero $\theta_{13}$ neutrino mixing angle has been measured~\cite{theta13}, observing or ruling out  neutrino CP violation becomes the {\em sine qua non} of neutrino physics, from which the needed neutrino factory performance follows. For this physics, the neutrino factory has been shown to be superior to all other facilities~\cite{IDS}.  A staged plan proceeding through a series of neutrino factories and muon colliders is under development~\cite{Palmer-etal-IPAC13,MASS-WP}, beginning with a short-baseline neutrino experiment ($\nu$STORM~\cite{NuSTORM}) employing a muon storage ring but without the need for muon cooling.\footnote{$\nu$STORM could also provide a development platform for subsequent muon technology tests.}

These advantages of muons are offset by technical challenges associated with the short muon lifetime and large beam size.  
Thus new, rapid, large-aperture beam manipulation, cooling, and acceleration techniques are required. 

\section{Muon Cooling}
The established beam-cooling methods are ineffective on the microsecond timescale of the muon lifetime. However,  the muon's penetrating character enables rapid  cooling via ionization energy loss~\cite{ionization-cooling,Neuffer-yellow}. The possibility that more novel techniques, such as the proposed optical stochastic cooling~\cite{Zh-Zo,Nagaitsev-IPAC12}\footnote{At sufficient energy for  appreciable muon synchrotron radiation in wigglers or undulators: $\sqrt{s}= 126$\,GeV (Higgs Factory), or $\sim$\,TeV.}, could meet the cooling-rate requirement is also under study.

\begin{figure}[t]
\includegraphics[width=\linewidth]{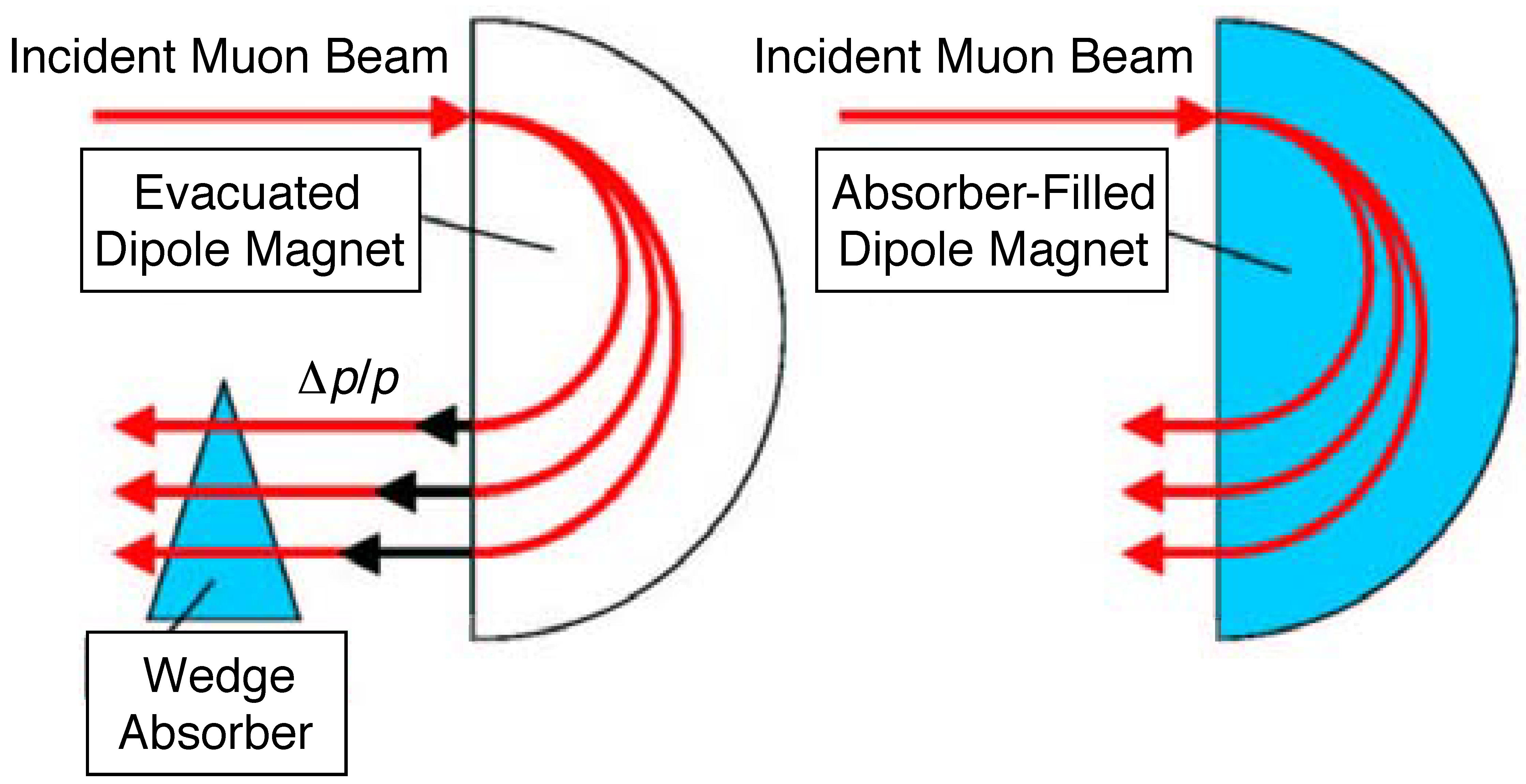}
\vspace{-.2in}\caption{Two approaches to emittance exchange: in each,
an initially small beam with nonzero momentum spread is
converted into a more monoenergetic beam with a spread
in transverse position. (Figure courtesy of Muons, Inc.).}\label{fig:emexch}
\end{figure}

An ionization-cooling channel comprises energy absorbers and radio-frequency (rf) accelerating cavities placed within a  focusing magnetic lattice. In the absorbers the muons lose both transverse and longitudinal momentum; the rf cavities restore the lost longitudinal momentum. In this way, the large initial divergence of the muon beam can be reduced.
Within a medium, normalized transverse emittance depends on path length $s$ as~\cite{Neuffer-yellow}
\begin{eqnarray*} 
\frac{d\epsilon_n}{ds}\approx
-\frac{1}{\beta^2} \left\langle\!\frac{dE_{\mu}}{ds}\!\!\right\rangle\frac{\epsilon_n}{E_{\mu}}
 +
\frac{1}{\beta^3} \frac{\beta_\perp
(0.014)^2}{2E_{\mu}m_{\mu}L_R},\ 
(1)
\label{eq:cool} 
\end{eqnarray*}  
where $\beta$ is the muon velocity in units of $c$,  $E_\mu$ the muon energy  in GeV, 
$m_\mu$
its mass in GeV/$c^2$, $\beta_\perp$ the lattice betatron function, and $L_R$ the radiation length of the medium. 
A portion of this cooling effect can be transferred to the longitudinal phase plane (``emittance exchange") by  
placing suitably shaped absorbers  
in dispersive regions of the lattice~\cite{Neuffer-yellow} or by using momentum-dependent path-length within a homogeneous absorber~\cite{HCC} (see Fig.~\ref{fig:emexch}). (Longitudinal ionization
cooling {\em per se}, which would entail operation at momenta above the minimum of the ionization curve so as to have negative feedback in energy, is impractical due to energy-loss straggling~\cite{Neuffer-yellow}).

\label{sec:matl}
The two terms of Eq.~\ref{eq:cool} 
 represent respectively muon cooling by  energy loss and heating by multiple Coulomb scattering. Setting them equal gives the equilibrium value of the
emittance, $\epsilon_{n,eq}$, at which the cooling rate reaches zero, and beyond which a given lattice cannot cool. Since the heating term scales with $\beta_\perp/L_R$, to achieve a low $\epsilon_{n,eq}$ requires
low  $\beta_\perp$ at the absorbers. Superconducting solenoids, which can give 
$\beta_\perp<\!\!<100$\,cm, are thus the focusing element of choice. Likewise, low-$Z$
absorber media are  
favored, the best being hydrogen (approximately twice as effective for cooling as the next best material, helium~\cite{Kaplan-COOL03}). 

\begin{figure}[bt]
\includegraphics[width=\linewidth]{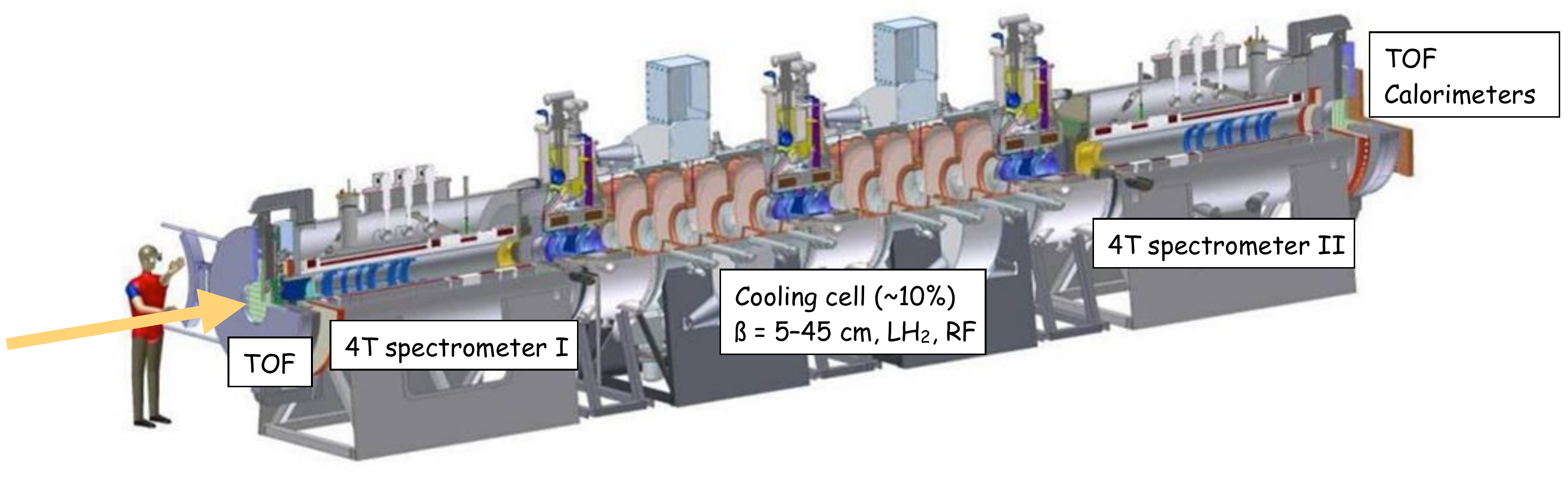}
\vspace{-.25in}\caption{Three-dimensional cutaway rendering of MICE apparatus (see text): individual muons entering at lower left are measured by time-of-flight (TOF) and (not shown) Cherenkov counters and a solenoidal tracking spectrometer; then, in cooling section, alternately slowed in LH$_2$ absorbers and reaccelerated by rf cavities, while focused by a lattice of superconducting solenoids; then remeasured by a second solenoidal tracking spectrometer, and their muon identity confirmed by TOF detectors and calorimeters.
}
\label{fig:MICE}
\end{figure}

It is the absorbers that cool the
beam, but for typical ``real-estate" accelerating gradients ($\approx$\,10\,MeV/m, to be compared with $\langle
dE_\mu/ds\rangle\approx30$ MeV/m for liquid hydrogen~\cite{PDG}), it is the rf cavities that  determine the
length of the cooling channel (see e.g.\ Fig.~\ref{fig:MICE}). 
The achievable rf gradient
thus determines how much cooling is practical before an appreciable fraction of the muons have
decayed. High-gradient vacuum rf cavities (normal-conducting due to the magnetic field in which they must operate) for muon cooling are under development, as is an alternative approach: cavities pressurized with hydrogen gas, thus combining energy absorption and reacceleration~\cite{RF-R&D}. In the first cooling stages the large size of the uncooled beam requires relatively low rf frequency. As the beam is cooled, focal lengths must be shortened in order to reduce the equilibrium emittance, and cavity frequencies and gradients can be increased. Goals are $\stackrel{>}{_\sim}$\,15\,MV/m at 201\,MHz in $\approx$\,2\,T fields, and $\approx$\,25\,MV/m at 805\,MHz in $\approx$\,3\,T. Promising results on meeting these goals are now coming from work at the Fermilab MuCool Test Area (MTA).

In the cooling term of Eq.~\ref{eq:cool}, 
the percentage decrease in normalized emittance is proportional to the percentage energy loss, thus (at 200\,MeV/$c$) cooling in one transverse dimension by a factor 1/$e$ requires $\sim$\,50\% energy loss and replacement.  Despite the relativistic increase of muon
lifetime with energy, ionization cooling thus favors low beam momentum,
due to the increase of
$dE/ds$ for momenta below the ionization minimum~\cite{PDG}, the greater ease of beam focusing, and the lower accelerating voltage required.  Most muon-cooling designs 
have used momenta in the range 150$-$400\,MeV/$c$. This is also the momentum range in which
the pion-production cross section from thick targets tends to peak and is thus optimal for muon
production as well as  cooling. The cooling channel of Fig.~\ref{fig:MICE}, for example, is optimized for a
mean muon momentum of 200\,MeV$/c$.  

\subsection{Towards a Muon Collider}

\begin{figure}
\includegraphics[width=.4\linewidth]{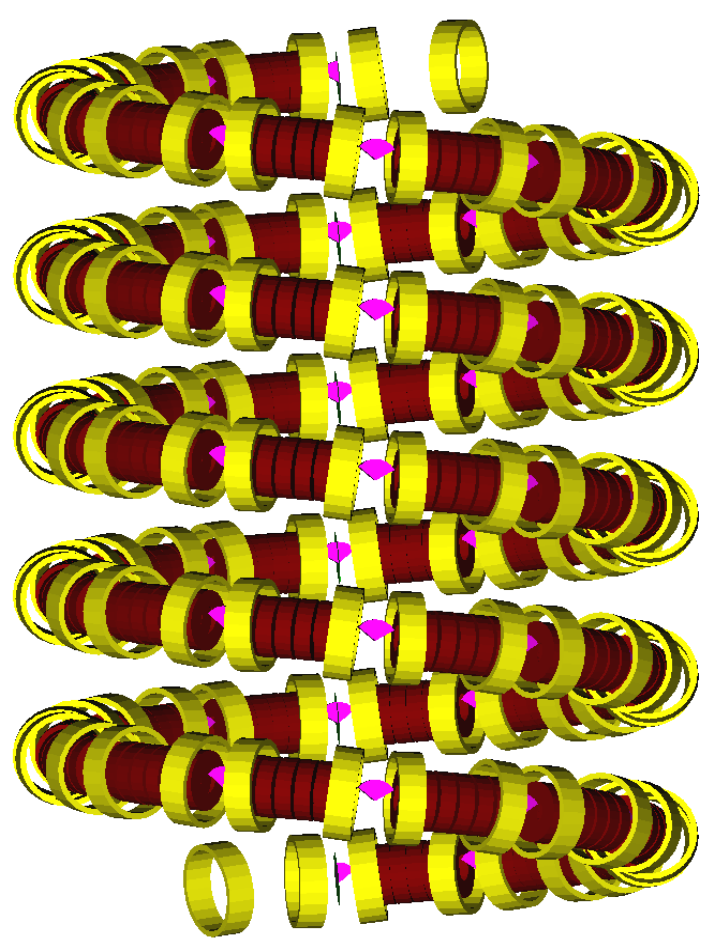}
\hbox{$\stackrel{\includegraphics[width=.6\linewidth]{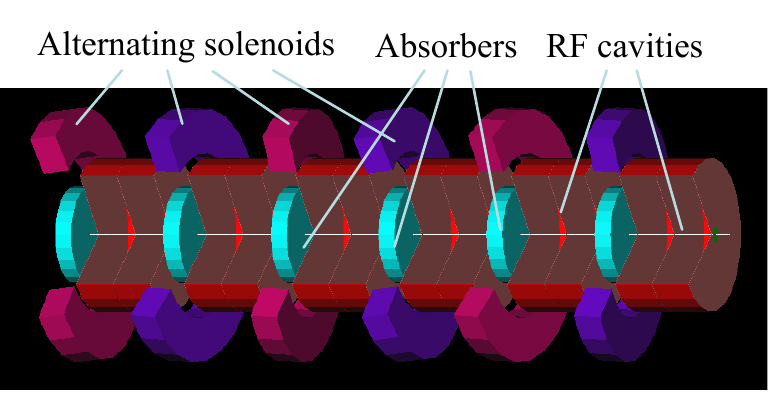}}{\includegraphics[width=.6\linewidth]{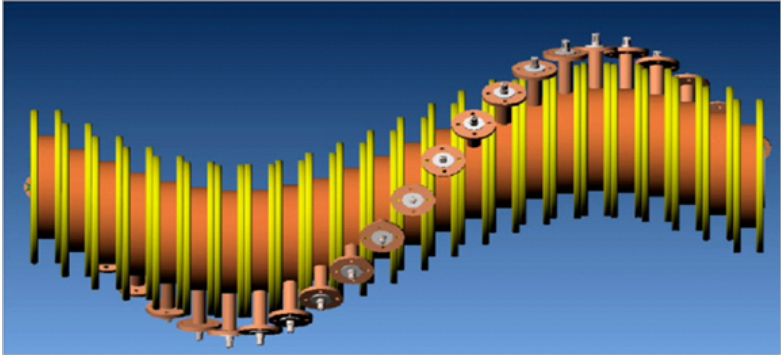}}$}
\vspace{-.2in}\caption{Three approaches to 6D cooling: (left) ``Guggenheim" helix~\protect\cite{Snopok},  (top right) ``FOFO Snake,"~\protect\cite{Alexahin} and (bottom right)  ``helical cooling channel" (HCC)~\protect\cite{HCC}.}\label{fig:6D}
\end{figure}

\begin{figure}
\includegraphics[width=\linewidth]{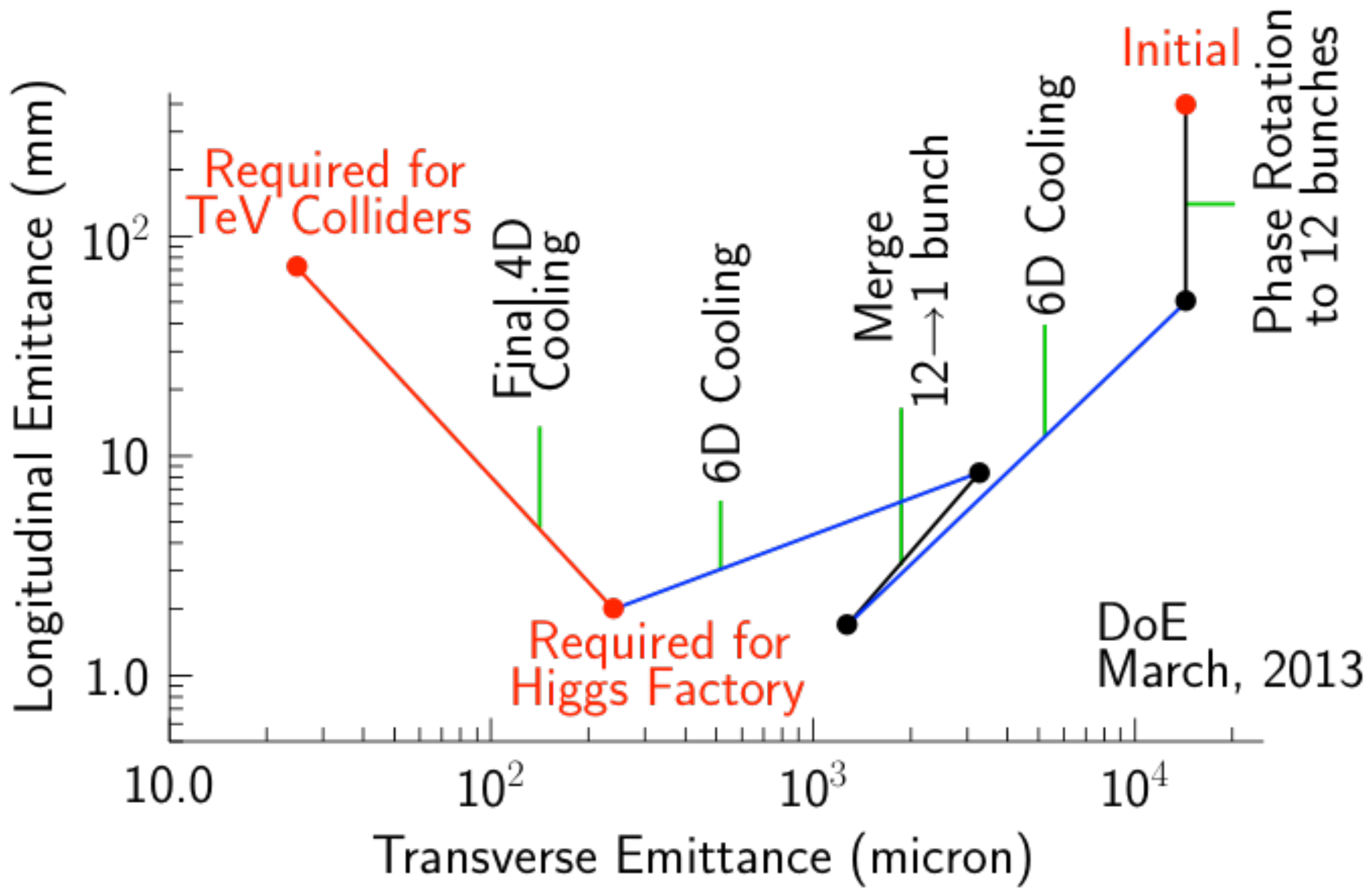}
\vspace{-.25in}\caption{Cooling trajectory in emittance space for Higgs Factory or multi-TeV muon colliders.}\label{fig:FN}
\end{figure}

Six-dimensional (6D) cooling lattices using longitudinal--transverse emittance exchange have received increasing attention in recent years~\cite{HCC,Palmer-complete,Snopok,Alexahin}. These are essential to a high-luminosity muon collider and may also enable  higher-performance or lower-cost neutrino factories. Three promising approaches are illustrated in Fig.~\ref{fig:6D}. All employ helical beam motion in order to create the dispersion needed for emittance exchange. The HCC~\cite{HCC} employs H$_2$-pressurized cavities while the baseline ``Guggenheim"~\cite{Snopok} and ``FOFO Snake"~\cite{Alexahin} designs use vacuum cavities. Figure~\ref{fig:FN} illustrates a possible trajectory in 6D emittance space that leads first to the optimal point for a Higgs Factory and then (via ``Final 4D Cooling"~\cite{Final}) to the optimum for a high-luminosity, multi-TeV collider. (The pieces of this trajectory have been simulated to demonstrate the needed performance, using both the Guggenheim and HCC approaches.)
The Higgs Factory optimum gives $\sim$\,0.003\% momentum spread at the IP, to match the narrow width of the Standard Model Higgs boson, whereas the $\sim\!\!10^{34}$\,cm$^{-2}$\,s$^{-1}$ luminosity needed at multi-TeV center-of-mass energy is achieved by further reducing the transverse emittance in the Final Cooling channel at the expense of increased longitudinal emittance. However, for both types of collider an overall 6D cooling factor of $\sim\!10^6$ is required.

The muon facility R\&D effort has identified a number of important new technologies for future muon facilities, for which a series of demonstration experiments are completed, in progress, or proposed:
(1)~the MERIT (Mercury Intense Target) experiment, carried out at CERN in 2007, showing feasibility of a mercury-jet target for a 4\,MW proton beam with solenoidal pion capture~\cite{MERIT};
(2)~EMMA (Electron Model of Muon Accelerator), a model ``non-scaling" fixed-field alternating-gradient (NS-FFAG) accelerator built and operated at Daresbury Laboratory~\cite{EMMA}\footnote{NS-FFAGs are not absolutely required for muon facilities but for certain muon acceleration stages may be more cost effective than RLAs.};
(3)~ MICE (the Muon Ionization Cooling Experiment), under construction at Rutherford Appleton Laboratory (RAL), aiming to verify the feasibility and performance of transverse ionization cooling by  2019~\cite{MICE,MICE-beam,MICE-emitt};
(4)~JEMMRLA, proposal for an electron model of a multipass-arc muon RLA~\cite{RLA} at Jefferson Laboratory~\cite{JEMMRLA}.
In addition, the Fermilab IOTA facility may soon be used to demonstrate optical stochastic cooling.

\section{MICE}

\begin{figure}
\hspace{.1in}\includegraphics[trim=20 200 10 110mm,clip,width=.9\linewidth]{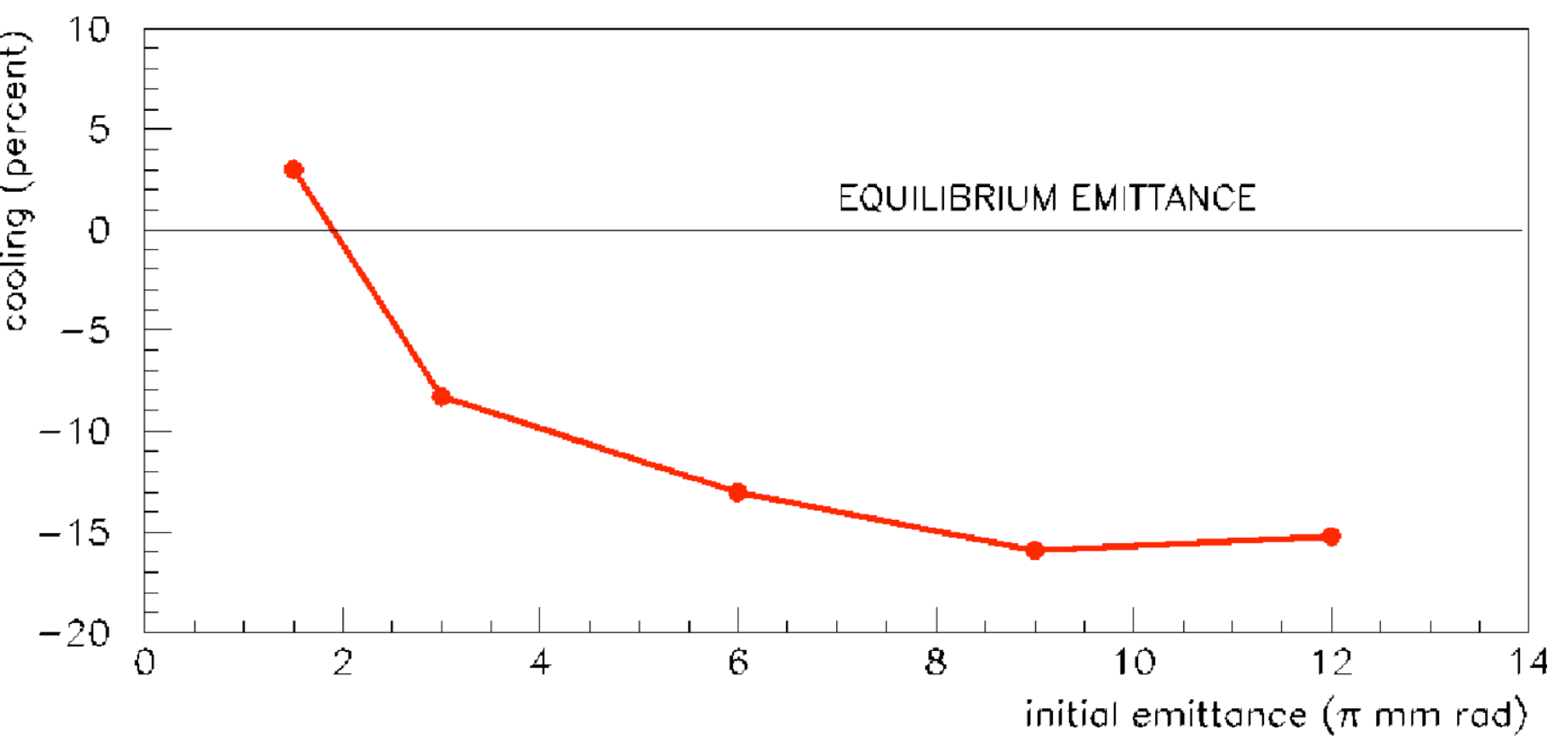}
\vspace{-.1in}\caption{Transverse emittance change in MICE cooling section vs input emittance for baseline MICE optics setting.}
\label{fig:perf}
\end{figure}

The Muon Ionization Cooling Experiment~\cite{MICE} seeks to demonstrate for the first time the feasibility and efficacy of ionization cooling of muons.
Figure~\ref{fig:MICE} shows the MICE apparatus: one cooling lattice cell (based on a design from Neutrino Factory Feasibility Study II~\cite{FS-II}) surrounded by the input and output spectrometers and particle-identification detectors that will be used to demonstrate and characterize the ionization-cooling process experimentally. Since an affordable cooling-channel section cools by only $\sim$\,10\% (Fig.~\ref{fig:perf}), too small an effect to measure reliably using standard beam instrumentation, MICE employs a low-intensity muon beam~\cite{MICE-beam} and measures each muon individually.
It will thereby demonstrate that the process is well understood in both its physics and engineering aspects, and works as simulated. 
In order to afford a thorough validation of the codes used to design ionization-cooling channels, MICE will be operated in a number modes and optics configurations.
The full results from MICE are expected by about 2020, with analyses of some configurations available some years earlier.
Early results are expected to include important validations of the models used in ionization-cooling simulation codes, as well as the first experimental test of muon transverse--longitudinal emittance exchange in a wedge absorber.

The MICE schedule has incurred significant delay due to the challenges of building the cryocooled magnet designs employed in the interest of MICE cost reduction. The first magnets are now complete: the two spectrometer solenoids (SS) and the first ``focus coil" (FC) solenoid pair used to  provide low beta at the absorber. Each SS cold mass comprises five separate coils: three to give a uniform 4\,T  ($\pm1$\%) field over the 1-m-long tracking volume and two to match into the cooling cell optics. Training and field-mapping of these magnets is in progress with the goal of first cooling measurements (but without rf reacceleration) in 2015. 

The acceleration sections include the most challenging MICE magnets: the two ``coupling coils" (CCs) surrounding the 201\,MHz rf cavities, each with stored energy of $\approx$\,6\,MJ. A first CC cold mass, for use in rf-cavity tests at the MTA,  has been built and is currently under test at the Fermilab Solenoid Test Facility. Once its performance is characterized, fabrication of the two MICE CCs will ensue. Given the complexity of the  ``RFCC" modules, their integration is necessarily a time-consuming task, with delivery to MICE planned in 2018. Integration of the entire MICE apparatus, with its coupled chain of superconducting magnets including a total of 18 coils operable in a variety of polarity and field settings, is a large and complex effort and will also take time.
A further challenge is magnetically shielding the MICE apparatus so that the surrounding electronic  and electrical equipment will work to specification. These are the issues driving the 2020 time frame for full MICE results.

\begin{figure}
\centerline{\includegraphics[width=.66\linewidth]{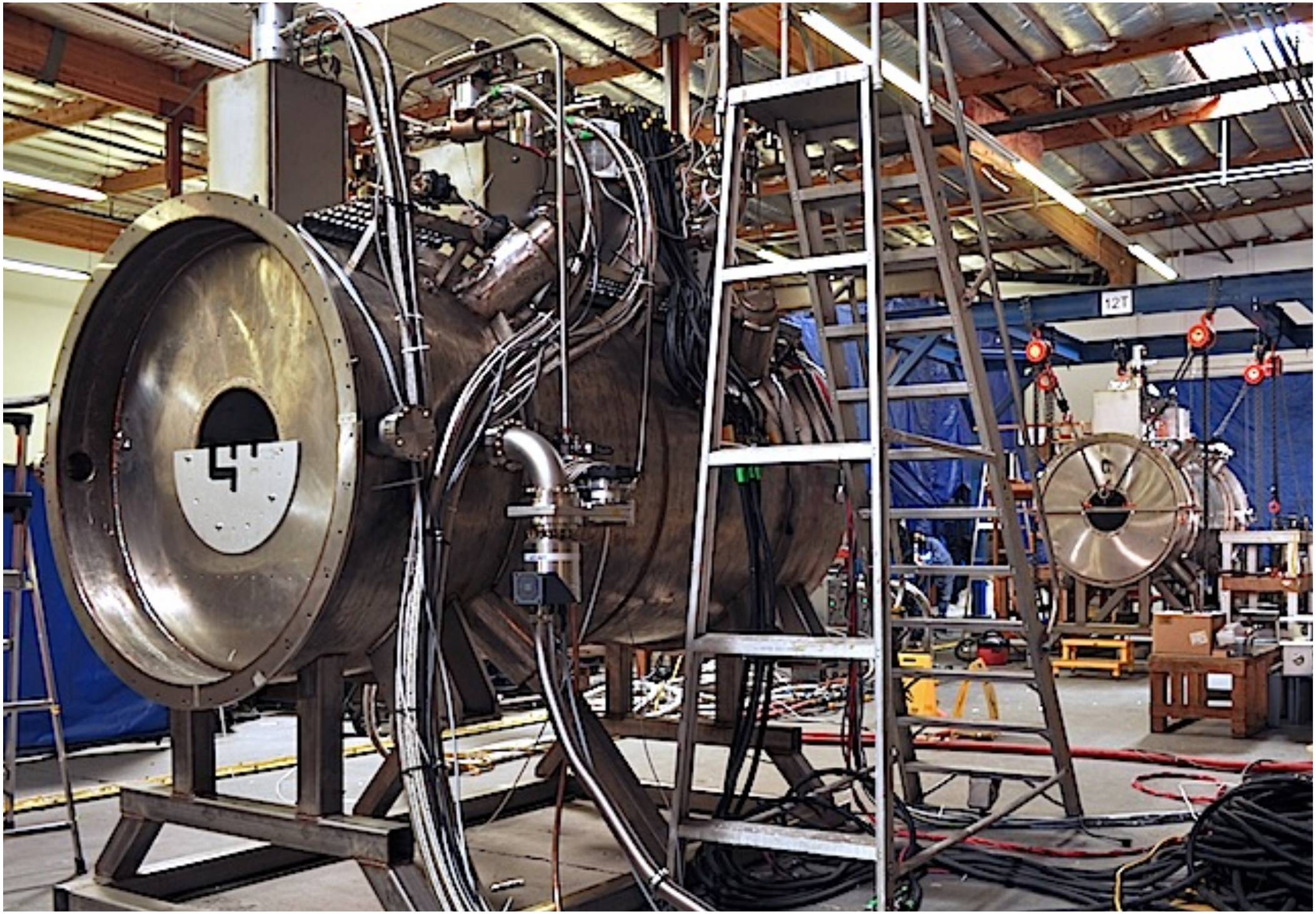}
\includegraphics[width=.28\linewidth,height=1.5in]{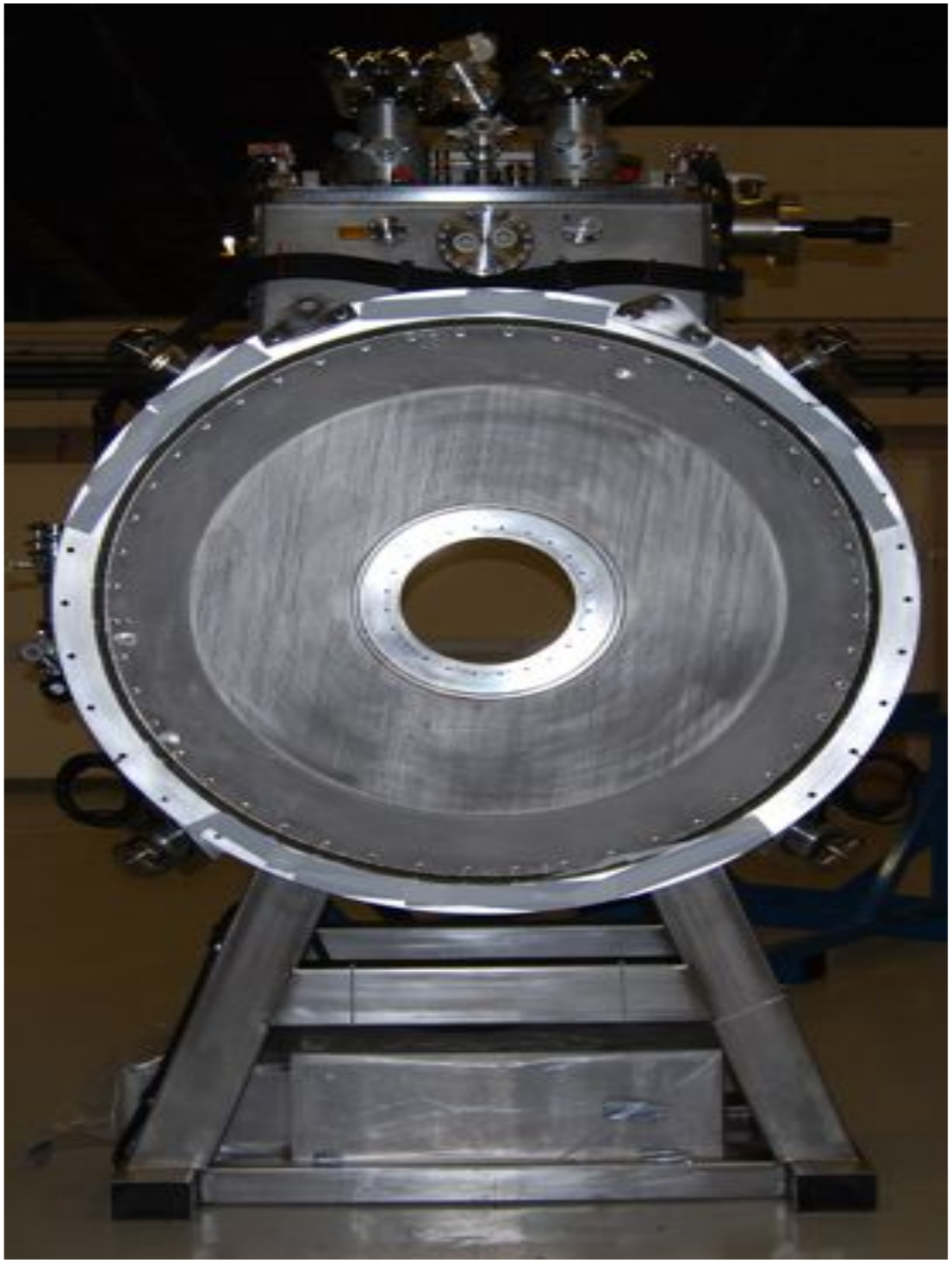}}
\caption{(left) MICE spectrometer solenoids at vendor; (right) first focus coil module under test at RAL.}
\label{fig:SS2}
\end{figure}

The MICE beam line and all detectors except the final ``electron--muon ranger" (EMR) calorimeter have been installed and their performance characterized~\cite{MICE-beam,MICE-emitt} (the scintillating-fiber trackers have been extensively tested with cosmic rays while awaiting the spectrometer solenoids). The EMR is in final assembly, with installation and beam tests  planned for later this year.

\section{Beyond MICE}

After MICE, additional technology demonstrations may be required. These are under study as part of the Muon Accelerator Program (MAP)~\cite{MAP}. Issues that may require experimental tests include collective effects of intense, low-energy beams in vacuum or traversing material. While no suitably intense muon beam is expected to be available prior to the first cooled-muon facility, analog experiments employing proton beams may be feasible.

\section{Conclusion}

Recent experimental results (discovery of the Higgs boson and nonzero $\theta_{13}$) have strengthened the physics case for a muon facility. With  key techniques established by $\approx$\,2020, construction of the first of the new generation of muon facilities could begin during that decade. 

\section{ACKNOWLEDGMENT}
The author thanks his MAP and MICE collaborators as well as the COOL'13 organizers for a most stimulating ride.


\end{document}